\documentclass[prb,aps,amsmath,amsfonts,amssymb,twocolumn, superscriptaddress,footinbib,showpacs, 10pt]{revtex4}
\usepackage{times,xspace}
\usepackage{amsfonts}
\usepackage{bm}
\usepackage{amssymb}
\usepackage[final]{graphicx}
\usepackage{color}
 
\begin{document}
 \title{Vortex-assisted photon counts and their magnetic field dependence in single-photon superconducting detectors}

\author{L. N. Bulaevskii}
\affiliation{Theoretical Division, Los Alamos National Laboratory, Los Alamos, New Mexico 87545, USA}

\author{Matthias J. Graf}
\affiliation{Theoretical Division, Los Alamos National Laboratory, Los Alamos, New Mexico 87545, USA}

\author{V. G. Kogan}
\affiliation{Ames National Laboratory, Ames, Iowa 50011, USA}
 
\date{January 12, 2012}
 
\begin{abstract}
We argue that photon counts in a superconducting nanowire single-photon detector (SNSPD) are caused by the transition from 
a current-biased metastable superconducting state  to the normal state. Such a transition is triggered by vortices  crossing the thin 
and narrow superconducting strip from one edge to another due to the Lorentz force. 
Detector counts in SNSPDs may be caused by three processes: 
(a) a single incident photon with sufficient energy to break enough Cooper pairs to create a normal-state belt
across the entire width of the strip (direct photon count),
(b)  thermally induced single-vortex crossing in the absence of photons (dark count), which at high-bias currents 
releases the energy sufficient to trigger the transition to the normal state in a belt across the whole width of the strip, and 
(c) a single incident photon of insufficient energy to create a normal-state belt   
but initiating a subsequent single-vortex crossing, which provides the rest of the energy needed 
to create the normal-state belt   (vortex-assisted single-photon count).  
We derive the current dependence of the rate of vortex-assisted photon counts. 
The resulting photon count rate  has a plateau at high currents  close to the critical current and drops  as a power-law with high exponent at lower currents. 
While the magnetic field   perpendicular to the  film plane does not affect 
the formation of hot spots by photons, it causes  the rate of vortex crossings (with or without photons) to increase.
We show that by applying a magnetic field 
one may characterize the energy barrier for vortex crossings and identify the origin of dark counts 
and vortex-assisted photon counts. 
\end{abstract}


\maketitle

\section{Introduction}

The superconducting nanowire single-photon detector (SNSPD) consists of a thin and long meandering superconducting strip carrying a bias current $I$ slightly below the critical current $I_c$,
where the superconducting state is metastable. When a photon interacts with the strip it creates a hot spot in 
the   film that drives a belt-like region across the width of the strip to the normal state. Consequently, a voltage pulse caused 
by the current redistribution between the superconducting strip and a  shunt resistor parallel to the strip is detected on nano-second time scales. 
After the  normal belt of the strip cools down,
 the strip returns into the metastable superconducting state. 
  Measuring voltage pulses,  single photons can be detected and counted. However, similar pulses are also recorded in the absence of photons, so-called dark counts, which introduce uncertainty in the counting of single photons.

Recently, thermally activated single-vortex crossing was proposed as a possible mechanism for dark counts in SNSPDs. \cite{bgbk}
As a consequence of the Lorentz force acting on a vortex crossing  a thin and narrow current-biased  strip the energy $\Phi_0 I/c$
is released, which for currents   $I \gtrsim  0.6 I_c$ suffices to create a normal belt 
across the entire width $w$ of the strip (extending to a few correlation lengths  $\xi$ along the strip). 
This process causes the transition from the current-carrying superconducting metastable state (S) 
of the strip at $I>I^*$  to the state with the normal (N) belt, and induces a current redistribution into the shunt 
accompanied by a measurable voltage pulse in the SNSPD. 
Here, $I_c$ is the critical current at which  
{\it the energy barrier vanishes for vortex crossing.}

It was shown in Ref.~\onlinecite{bgbk} that in a thin, $d\alt \xi \ll \lambda$, and narrow, $w\ll \Lambda$, film
 the single-vortex crossing has the lowest energy barrier for creating
dissipation, while  phase slips across the entire width of strip and vortex-antivortex depairing are 
characterized by higher barriers. Here $\Lambda=2\lambda^2/d$ is the Pearl length with London penetration depth $\lambda$
and film thickness $d$.
In fact, calculations accounting for the mechanism of vortex-antivortex binding and unbinding
resulted in an energy barrier at least twice as large as for a single-vortex crossing.\cite{bgbk, Sem}
 Since all these processes are thermally activated, that is,  their rates depend exponentially on the barrier height, 
one can safely consider only the process with the lowest barrier. 
The vortex-crossing rate derived in Ref.~\onlinecite{bgbk} 
as  a function of $I$, $w$ and temperature $T$ is in  agreement with   dark count rates measured by  Bartolf {\it et al.} \cite{Bartolf}
Recently,  Hofherr {\it et al.}\cite{Semlast}  
also ascribed the single-vortex crossing mechanism to 
the origin of dark counts  in SNSPD.

Here, we show that a weak magnetic field 
perpendicular to the superconducting strip results 
in a significant suppression of the energy barrier for vortex crossing. This leads to an increase in 
the vortex-crossing  rate. 
Hence, the vortex-based mechanism for dark counts can be tested experimentally. 
Also, we  consider  the case when the energy of a single photon is not large enough to create the  normal  belt across the entire strip  width. 
Rather it leads to a belt-like superconducting region with an elevated quasiparticle temperature.
In this situation a photon-induced hot spot enhances strongly 
the probability of a subsequent vortex 
crossing. The combined effect of hot spot and vortex crossing leads to the creation of a normal 
belt across the strip, i.e., to a single photon count. 
We call such photon counts ``vortex-assisted''.

In  recent SNSPD experiments, Hofherr {\it et al.} \cite{Semlast} showed for a NbN strip, 97 nm wide, that photons of  500\,nm wavelength 
give rise to a photon count rate that
decreases at low currents. Similar results were obtained by the NIST group \cite{Baek} 
for a-W$_x$Si$_{1-x}$ based SNSPD.
Our interpretation of these experiments fits the situation described above, i.e.,
a single photon cannot turn an entire belt into the N state
 and additional energy due to vortex crossing is needed for an S-N transition to happen in the belt.
As the probability for vortex crossing depends 
strongly on the bias current and drops significantly at currents well below $I_c$, \cite{bgbk,gv} 
the photon count rate becomes current dependent at low currents.  

\begin{figure}[t]
\includegraphics[width=80mm]{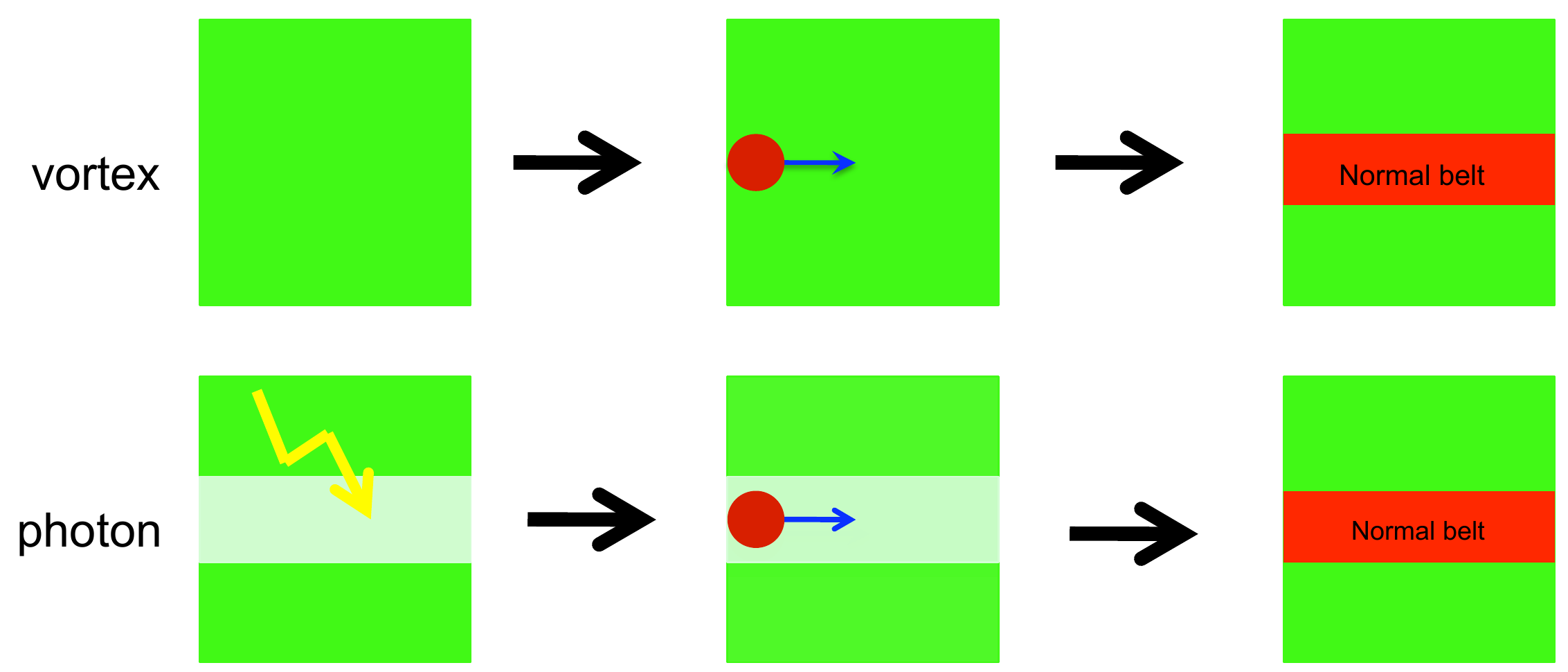}
\caption{(Color online)
Cartoon   of a superconducting strip with a vortex crossing without and with photon.
Top: from left to right, illustration of the thermally excited   vortex crossing and subsequent formation of a normal-state hot belt across the strip width resulting in a dark count.
Bottom:    an incident photon creates a hot spot (hot belt) across the superconducting strip, followed by
the thermally induced vortex crossing, which turns superconducting hot belt into the normal state resulting in a vortex-assisted photon count.}
\label{fig:cartoon}
\end{figure}

Hence, our general picture is that a   vortex crossing 
may trigger the S$\rightarrow$N transition. A photon  makes this process much more probable
by creating a  spot with suppressed superconducting order parameter and thus with lower energy barrier for vortex crossing.
A sketch of the   strip and of the belt  across are shown in Fig.\,\ref{fig:cartoon}.
Like the bias current,  an applied magnetic field enhances the rate of  vortex 
crossings and causes an increase of the
photon count rate. Therefore, the mechanism of vortex-assisted photon counts can be verified 
experimentally because the magnetic field does not affect the creation of hot spots,   
but increases the probability of vortex crossings.

\section{Effect of the magnetic field on the vortex-crossing rate}

\subsection{Basic equations}

In the following ${\bf r}=(x,y)$ is a position on the thin  strip,
 $0 \leq x \leq w$,  $-L/2 \leq y \leq L/2$, and   $w\ll L$. In the presence of a vortex at position  ${\bm r}_v$,
 the superconducting order parameter, normalized to its   uniform zero-current  value, 
reads
\begin{equation}
\Psi({\bm r}, {\bm r}_v)=\mu\exp\{i[\varphi({\bm r}, {\bm r}_v)-\kappa y/\xi+\varphi_0]\},
 \label{Psi}
\end{equation}
 see, e.g., Refs.\,\onlinecite{Amb, McC}. 
Here $\mu^2=1-\kappa^2$  describes the suppression of the  order parameter by the bias current. In Eq.\,(\ref{Psi}), $|\Psi|$ is a constant, i.e.,  its behavior in the vortex core is disregarded. Hence, we employ a variation of the London approach in which, however, the Ginzburg-Landau suppression of the order   parameter by transport supercurrents is taken into account.
The parameter $\mu$ does not change much 
 when the current varies from 0 to $I_c$, because  $1 < 1/\mu^2 < 1.157$.
The critical current $I_c$ is defined by a vanishing energy barrier, 
  see, e.g., Ref.\,\onlinecite{bgbk}, and is given by 
\begin{equation}
I_c\approx\frac{c\Phi_0w}{4\pi^2e\tilde\Lambda\xi} \approx \frac{2w}{\pi e\xi}I_0 \,,\quad I_0=\frac{c\Phi_0}{8\pi\tilde{\Lambda}},  
\label{eq:Ic}
\end{equation}
 $e\simeq 2.718$. In this expression the parameter $\xi$ originates from the cutoff of the supercurrent at distances of the superconducting correlation length in the London approach. 
Hence, the critical current, Eq.~(\ref{eq:Ic}), determined in the London approach has an uncertain numerical 
factor of order unity. Note, that the critical current depends on the film configuration, for more details 
see Ref.~\onlinecite{CB}. 
It is easy to see that this current is of the same order as the depairing critical current  density $c\Phi_0/12\sqrt{3}\pi^2 \lambda^2\xi$ 
within Ginzburg-Landau theory, when multiplied by the strip cross-section $wd$.  
Since $|\Psi|$ is suppressed by the supercurrent $I$,   the Pearl length should be renormalized accordingly $\tilde\Lambda=\Lambda/\mu^2$.  

In   narrow strips, $w\ll \tilde{\Lambda}$,  supercurrents are found by neglecting the induced magnetic field   and the corresponding vector potential.\cite{bgbk}
{Then the sheet current density ${\bf g}$ for the vortex at position ${\bm r}_v=(x_v,0)$ is ${\bm g}={\rm curl}\, G\hat{{\bm z}}$,
\begin{equation}
 G({\bm r})=\frac{I_0}{\pi}\ln\frac{\cosh Y-\cos(X+X_v)}
{\cosh Y-\cos(X-X_v)},\quad \qquad
\end{equation}
where $X=\pi x/w$ and $Y=\pi y/w$ are dimensionless coordinates.  
{The superconducting order parameter phase is\cite{bgbk}
\begin{equation}
\varphi({\bf r})=\tan^{-1}\frac{\sin X_v\sinh Y}{\cos X-\cosh Y\cos X_v}.
\end{equation}
Note that the characteristic length of the sheet current and phase variations  for noninteracting vortices in both $x$ and $y$ directions is $w$. \cite{bgbk}
In a strip of length $L\gg w$, a single vortex far away from the strip ends  
creates the phase difference 
\begin{equation}
\varphi(L/2)-\varphi(-L/2)=2X_v ,
\end{equation}
so that a vortex crossing from  $X_v=0$ to $X_v=\pi$ changes the phase difference by $2\pi$. This is a generalization 
of phase slips in one-dimensional (1D) wires. 
The vortex crossing from one edge to another then causes the phase difference at the strip ends to vary in time.
This induces a voltage change along the length $L$ of the strip. The Josephson equation relates voltage and phase, which in our case relates the induced voltage with vortex velocity: \cite{Clem} 
\begin{equation}\label{JosephsonRelation}
V(t)=\frac{\Phi_0}{2\pi c}\frac{d}{dt}[\varphi(L/2)-\varphi(-L/2)]=\frac{\Phi_0}{cw}\frac{dx_v}{dt}. 
\end{equation}

The energy of the vortex in zero applied magnetic field and at zero bias current is   $(\Phi_0/2c)G(x\to x_v,0)$, whereas 
in the presence of the bias current $I$ one needs to add the term $\Phi_0 IX_v/\pi$ due to the Lorentz force. 
In the applied   field ${\bm H}=H {\bm z}$, we also need to add the magnetic term 
$-M_v H$, where $M_v$ is the magnetic moment of the vortex:\cite{kogan}
\begin{equation}
M_v=\frac{1}{2c}\int d{\bm r}\, {\bm r}\times {\bm g}=\frac{\Phi_0}{4\pi\tilde{\Lambda} }x_v(w-x_v).
\end{equation}
Finally, the potential energy $U$ of a vortex in the presence of the bias current $I$ and the external  field $H$ is given by:
\begin{eqnarray}
\epsilon_0^{-1} { U(I,H,X_v) } &=&
 \ln\left(\frac{2w}{\pi\xi}\sin X_v  \right)
 -\frac{ I }{ I_0 } X_v
 \nonumber\\
&-&\frac{ H }{ H_0 } X_v \left( 1-\frac{X_v}{\pi} \right).
\label{energy}
\end{eqnarray}
Here $\epsilon_0=\Phi_0I_0/ \pi c =\Phi_0^2/ 8\pi^2\tilde{\Lambda} $
is the characteristic vortex energy in a thin film and 
$H_0 = \Phi_0/2w^2  $. 
The  potential energy  in Eq.~(\ref{energy}) corresponds to the vortex aligned with the applied field in agreement with 
Refs.~\onlinecite{kogan, Stejic1994}. 
However, it differs by a factor of 2 in the logarithm with the potential employed  by Gurevich and Vinokur.\cite{gv}
Note that the equation\,(\ref{energy}) for the potential energy 
is valid only when the vortex core is far enough from the edges. 

In the following, we consider the situation of magnetic fields $H<H_{c1}(w)$, where $H_{c1}(w)$ is the  field that corresponds to the  vortex at $X_v=\pi/2$ (the middle of the strip).
Thus the lower critical field for vortex entry is determined by $U(0,H_{c1}, \pi/2) = 0$ and yields
\begin{equation}
H_{c1}(w)=\frac{4H_0}{\pi} \ln\frac{2w}{\pi\xi}\,.
\label{eq:Hc1}
\end{equation} 
In Figs.~\ref{fig:potential_h0} through \ref{fig:potential_h4} we present the vortex energy as a function of position on the strip for different bias currents and magnetic fields.

\begin{figure}[t]
\includegraphics[width=65mm]{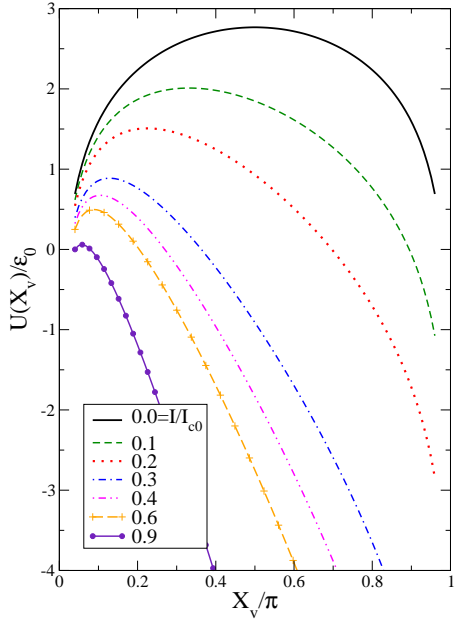}
\caption{(Color online)
Position dependence, $X_v = \pi x_v/w$, of the potential energy for vortex entry in zero magnetic field for several bias currents.
The potential is cut off for distances less than one coherence length $\xi$ from the left and right edges of  strip, because of the finite vortex core.
}
\label{fig:potential_h0}
\end{figure}

\begin{figure}[t]
\includegraphics[width=65mm]{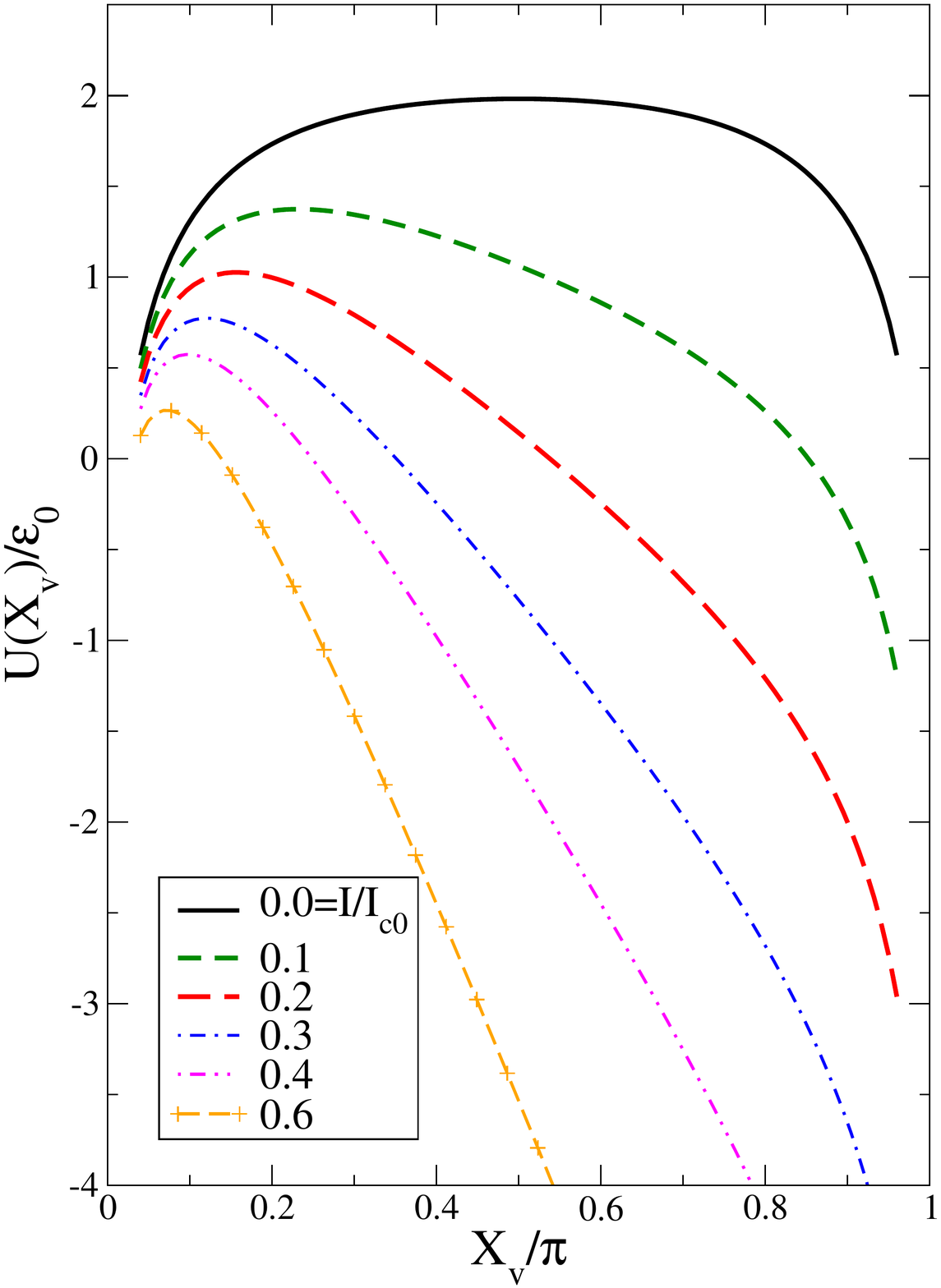}
\caption{(Color online)
Position dependence of the potential energy for vortex entry for $ H=H_0 $ for several bias currents.
}
\label{fig:potential_h1}
\end{figure}

\begin{figure}[t]
\includegraphics[width=65mm]{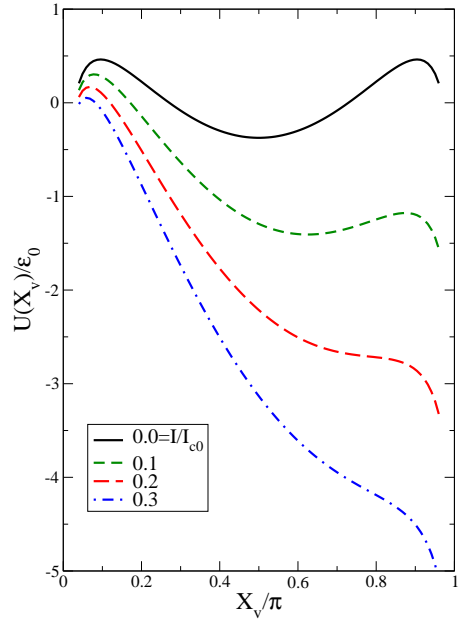}
\caption{(Color online)
Position dependence of the potential energy for vortex entry  for $ H=4H_0 $ for several bias currents.
}
\label{fig:potential_h4}
\end{figure}

\subsection{Barrier for vortex crossing in magnetic field}

\begin{figure}[tbh]
\includegraphics[width=65mm]{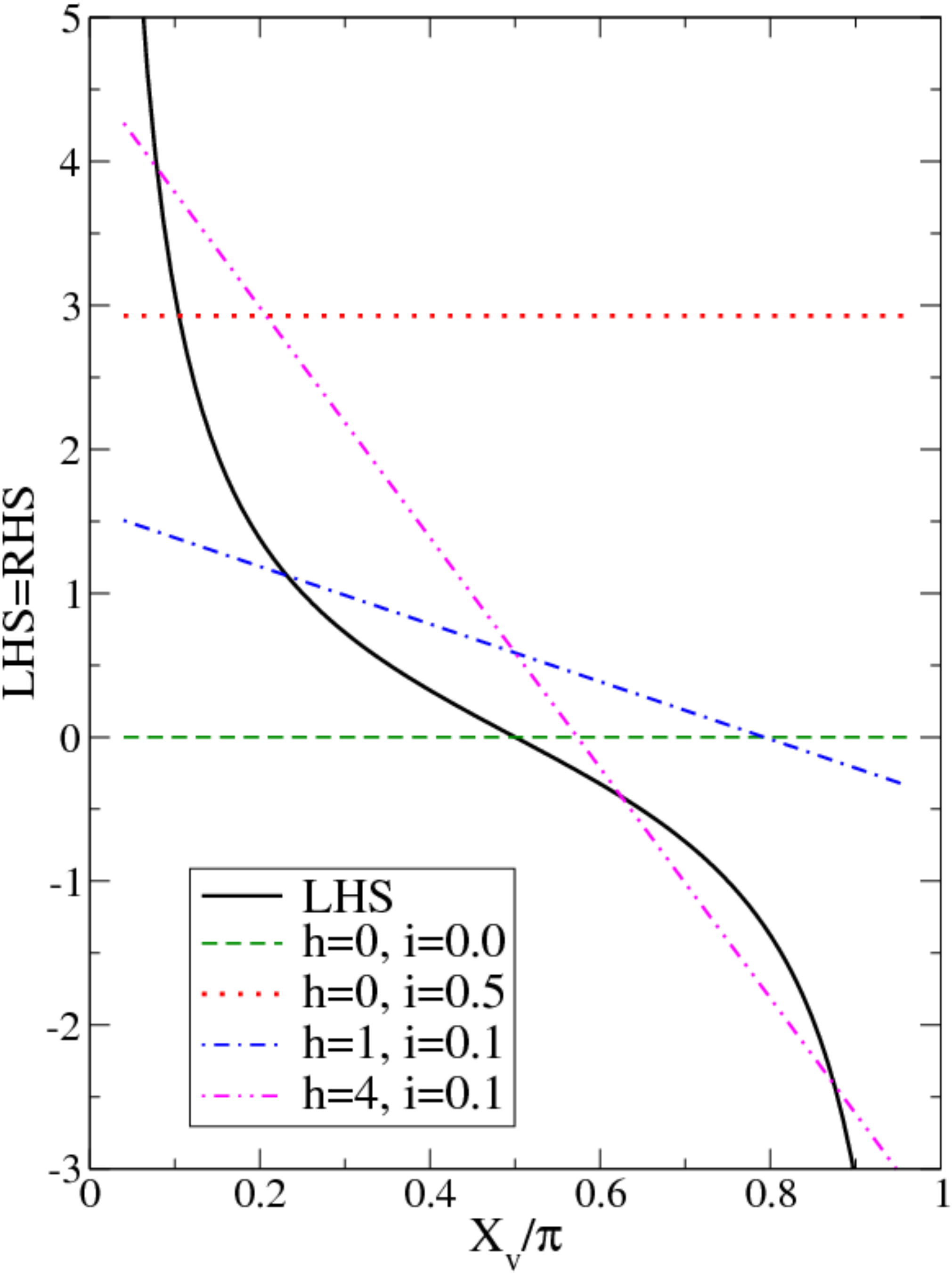}\caption{(Color online)
Graphic  solutions for barrier heights for several currents $i\equiv I/I_{c0}=0, 0.1, 0.5$ and fields $h\equiv H/H_0=0, 1, 4$ obtained by equating the LHS and RHS of Eq.~(\ref{cot}).
}
\label{fig:graphical}
\end{figure}

The energy barrier $\epsilon_b(I,H)$ is given by the maximum of 
$U(x_v)$   in the interval $\xi<x_v<w-\xi$, where the expression (\ref{energy}) holds
(For short we dropped the dependence of $U$ on $I$ and  $H$ and use the coordinates $x_v$ and $X_v$ interchangeably).
The positions of the extrema of $U(x_v)$   are determined by  $dU(X_v)/d X_v=0$:
\begin{equation}
{{\rm cot}\, X_v} =\frac{I}{I_0}+\frac{ H }{H_0}\left(1-\frac{2}{\pi}X_v\right) .
\label{cot}
\end{equation}
In Fig.\,\ref{fig:graphical}, we plot separately the left-  and right-hand sides (LHS and RHS) 
of this equation
for different bias currents $I$ and fields $H$ 
 to show qualitatively the solutions of Eq.\,(\ref{cot}).
For $H<\pi  H_0/2$ only one intersection exists
at $X_v=X_s$ 
corresponding to the barrier energy $\epsilon_b$. For fields $H>\pi H_0/2$  and small currents one finds three intersections 
corresponding to a global maximum, local minimum, and local maximum of $U(x_v)$, respectively,
see Figs.\,\ref{fig:potential_h4} and \ref{fig:graphical}.
On the other hand, for large currents and large fields only one extremum (maximum) exists.
At $I=0$ a local minimum exists in the middle of the strip at $X_v=\pi/2$ for fields $H>3H_0/2$.
However the vortex should overcome the barrier to reach this local minimum. The saddle point position
for this barrier is shown in the top panel of Fig.~\ref{fig:barrier}, while its suppression with magnetic field is shown in the bottom panel.

In the following, we discuss the case of   $I\gg I_0$ as $I_0$ is rather small in comparison with the currents on the order of $I_c$
employed usually  in SNSPD experiments ($I_c\sim (w/\xi)I_0 \gg I_0$).
The position of the saddle point (global maximum) 
$X_s\ll 1$ 
is given by 
\begin{equation}
\frac{1}{X_s}\approx 
\frac{I}{I_0}+\frac{H}{H_0} , 
\end{equation} 
when $X_s>\pi \xi/w$. Thus the energy barrier $\epsilon_b(I,H)$, see also Fig.~\ref{fig:barrier}, for vortex entry is given by
\begin{equation}
\frac{\epsilon_b}{\epsilon_0} \approx
\ln \frac{2w I_0/(\pi e\xi )}{I+I_0 H / H_0} =
\ln \frac{I_{c0}}{I_{+}} \, ,
\label{barrier}
\end{equation}
where $I_{c0}$ is the zero-field critical current given by Eq.~(2).  
We see that for large currents, one can account for the  field effects by renormalizing the current:  $I_{+}=I+I_0H/H_0$.  Therefore, at high currents, the vortex-crossing rate $R_v(I,H,T)$ can be obtained from the zero-field rate $R_v(I,T)$  
by replacing $I \to  I_{+}$. This approximation holds for large currents not too close to $I_{c0}$ to fulfill the condition $X_s>\pi \xi/w$.
From Eq.~(\ref{barrier}) it follows that for sufficiently small magnetic 
fields, $H \ll H^*$,
the field dependence of the critical current is linear in $H$,
\begin{equation} 
I_c(H)=I_{c0} \left(1-\frac{H}{H^*}\right), \ \ \ H^*\approx\frac{2 w}{\pi e \xi} H_0 .
\label{eq:H*}
\end{equation}

\begin{figure}[t]
\bigskip\bigskip
\includegraphics[angle=0,width=85mm]{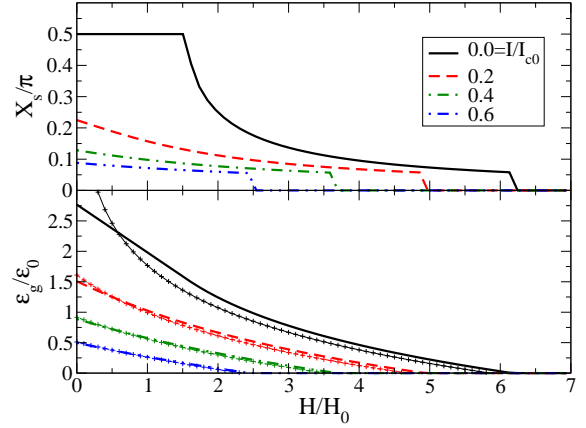}\caption{(Color online)
The position $X_s$ of the maximum of the vortex potential $U(x)$ (top) and the barrier height $\epsilon_b$   
(bottom) as  functions of applied 
perpendicular magnetic field for different bias currents.
The thin lines with ``$+$" symbols are the corresponding approximate barriers from Eq.~(\ref{barrier}),
which are in excellent agreement with the exact solutions for larger currents.
}
\label{fig:barrier}
\end{figure}

\subsection{Vortex-crossing rate}

{In Refs.\,\onlinecite{gv,bgbk}, the vortex-crossing rate was obtained by employing the known stationary solution for a potential $U(x_v)$
with periodic boundary conditions (i.e., for the potential $U(x_v)$ created by periodic extension to the whole axis $-\infty<x_v<+\infty$]. Here, we derive the vortex-crossing rate for the realistic potential 
$\tilde{U}(x_v)$, which is not periodic. 
In order to correctly treat vortex-assisted photon counts, we need a time-dependent  solution. 
We will use the rate derived in the quasi-stationary approach 
to find vortex crossing rate via relaxing hot spot and we will present condition when such an approach is valid. 
Besides, we show that the dissipation in the shunt resistor is comparable to the Bardeen-Stephen dissipation in the vortex core and cannot be neglected 
considering the dissipative vortex dynamics in SNSPDs. 

\subsubsection{Langevin equation}
The Bardeen-Stephen dissipation power is   $\eta_v\dot{x}_v^2$, where 
$\eta_v=\Phi_0^2/(2\pi \xi^2c^2R_{\square})$ is the Bardeen-Stephen drag coefficient.
Here $R_{\square}$ is the film sheet resistance   taken slightly above $T_c$. 
The power dissipated in the shunt with resistance $R_s$, 
due to the voltage pulse of a crossing vortex, is 
$V(t)^2/R_s= (\Phi_0\dot{x}_v/cw)^2/R_s \equiv \eta_s \dot{x}_v^2$, 
see Eq.~(\ref{JosephsonRelation}) for voltage-phase relationship.
Therefore the equation of vortex motion in the presence of shunt and thermal noise is
\begin{equation}
\eta_{\rm eff}\dot{x}_v=-\frac{\partial U(x_v)}{\partial x_v}+F_{L}(t),
\end{equation}
where $F_L$ is the Langevin force with the effective drag coefficient
\begin{equation}
\eta_{\rm eff}\equiv \eta_v + \eta_s=\frac{\Phi_0^2}{c^2}\left(\frac{1}{2\pi\xi^2R_{\square}}+\frac{1}{w^2R_s}\right).
\end{equation}
For the parameters $w=100$ nm, $\xi=4$ nm, $d=6$ nm, $R_{\square}=400$ $\Omega$ and $R_s=9$ $\Omega$ the shunt contribution to the viscosity 
is $\approx 0.45 \eta_v$ and thus cannot be neglected. 

\subsubsection{Fokker-Planck equation}
To find the vortex-crossing rate for diffusive vortex motion, 
we use the Fokker-Planck equation and Kramers' procedure to calculate 
the escape rate for particle diffusion over a barrier as described in Ref.~\onlinecite{Risken}. 
As mentioned previously, in the London approximation, one can treat  the vortex as a particle 
described by the potential energy in Eq.~(\ref{energy}) for coordinates $x_0<x_v<w-x_0$.
At the left edge, i.e.,  at distances smaller than $x_0\sim \xi$ an excitation precedes the crossing of the vortex (so-called ``pre-vortex''). 
In the presence of the bias current with its self-induced magnetic field lines at the edges,
the pre-vortex has a normal core, but no circular supercurrents enclosing it. 
Therefore, within the London approximation the boundary condition on the left side may be recast in the form of a trapped particle (pre-vortex) in a box potential sketched in Fig.~\ref{fig:FPBC} with the effective potential 
\begin{eqnarray}
\tilde{U}(x_v)=\left\{ 
\begin{array}{r@{\quad:\quad}l} 
\infty & x_v=0, \\  0 & 0<x_v\leq x_0, \\ U(x_v) & x_v > x_0.
\end{array} 
\right.
\end{eqnarray}
The infinite repulsive potential $\tilde U$ at $x_v=0$ prevents the pre-vortex from escaping to the left, because pre-vortex or vortex (particle) exist only
inside the superconducting strip.
The probability density of such particle is $W(x_v)\approx x_0^{-1}\exp[-\tilde{U}(x_v)/T]\approx  x_0^{-1}$ 
between $0 < x_v \leq x_0$, where we take into account the uncertainty of the vortex coordinate of the order $x_0$.
At $x_v>x_0$ the vortex is fully developed inside the strip and its diffusion over the barrier 
between points $x_0$ and $w-x_0$ is described by the potential $U(x_v)$.

\begin{figure}[bh]
\bigskip\bigskip
\includegraphics[angle=0,width=85mm]{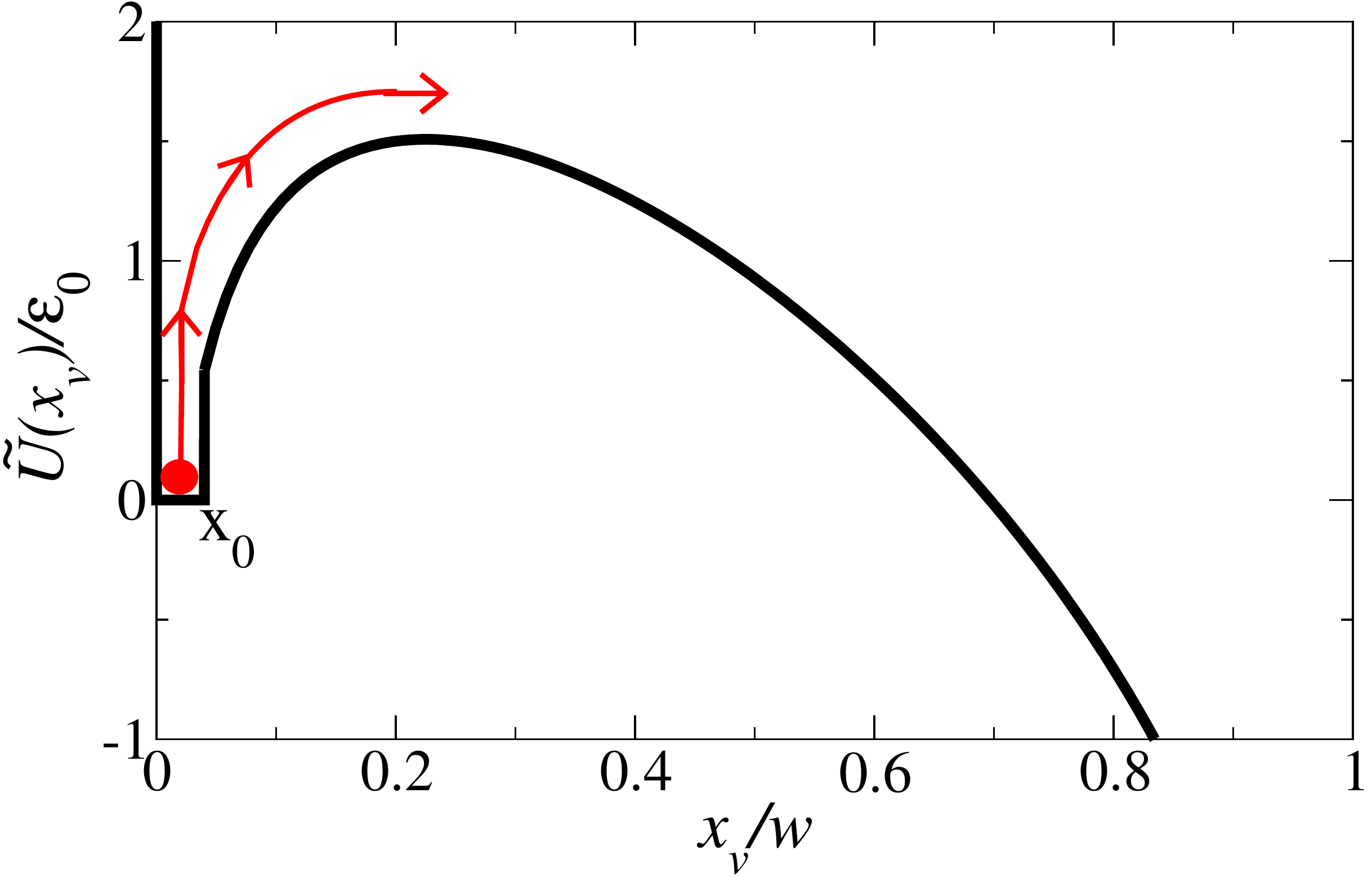}\caption{(Color online)
The effective potential $\tilde U(x_v)$ in the presence of bias current for a vortex in the particle approximation (red circle).
The left-side boundary condition of the Fokker-Planck equation describes the existence of a ``pre-vortex'' 
within distance $x_0\sim\xi$ of the strip's left edge.
}
\label{fig:FPBC}
\end{figure}

Finally, we solve the Smoluchowski (Fokker-Planck) equation for 
 a weakly time-dependent potential $\tilde{U}(x_v,t)$ to account for changes of the superconducting order parameter with time due to 
the hot spot relaxation:
\begin{eqnarray}
&&{\partial_t W(x,t)}=-{\partial_{x} S(x,t)}, \\
&&S(x,t)=-De^{-\tilde{U}(x,t)/T} {\partial_{x}} \left[
e^{\tilde{U}(x,t)/T}W(x,t) \right],
\label{Sm}
\end{eqnarray}
where $D=T/\eta_{\rm eff}$ is the diffusion coefficient,  $W(x,t)\,dx$ is the 
probability to find the vortex in the interval $(x,x+dx)$ 
at the time $t$, while $S(x,t)$ is the probability current density.
We consider bias currents $I<I_c$, such that $\epsilon_b/T$ is large and the crossing rate is small. Then $S$ and 
 $\partial_t W$ are  small, while the probability current density $S(t)\equiv S(x,t)$ is approximately independent of $x$, because $\partial_{x} S \approx 0$.
At the left edge, the probability density of a vortex is $W(x_0,0)=1/x_0$.
At the right edge, $W(w-x_0,t)$ is exponentially small as the
vortex escapes and $W(w-x_0,t)=0$ can be taken there. 
Integrating Eq.\,(\ref{Sm}) between $x_v=x_0$ and $x_v=w$, we obtain for the current probability
\begin{equation}
(D/x_0 )W(x_0,t)=S(t)\int_{x_0}^w dx \, e^{{U}(x,t)/T}.
\end{equation} 
Our solution for $S(t)^{-1}$ coincides with the one found by Kramers's escape rate for the double-well potential, see (5.109) in Ref.~\onlinecite{Risken}, if we set 
the potential $f(x)=0$ in the interval $(x_1,x_2)$  and take $p=1$ in (5.109).  
Also neglecting the exponentially small probability for vortex crossing from the opposite side, that is, from $x_v=w$ to $x_v=0$, 
we find the vortex-crossing rate 
\begin{equation} 
R_v(t) \equiv S(t)\approx \frac{D}{x_0}\left[\int_{x_0}^{w} dx \, e^{U(x,t)/T}\right]^{-1}.
\label{Rv}\end{equation}
Note that by symmetry the rate of antivortex crossing from $x_v=w$ to $x_v=0$ is the same as for vortex from
$x_v=0$ to $x_v=w$.
The main contribution to the integral in (\ref{Rv}) comes from regions near the maximum of $\tilde{U}(x)$, where it 
coincides with $U(x)$ for currents not too close to the critical one. 
Under this condition we can make the approximation
\begin{eqnarray}
R_v^{-1}D &\approx& x_0 \int_{0}^w dx \, e^{U(x,t)/T} 
\end{eqnarray}
Further, we can replace the prefactor
$x_0$ with $\xi$ to obtain 
\begin{eqnarray}
R_v^{-1}D &\approx& \frac{\xi^2}{2} \left( \frac{2w}{\pi\xi} \right)^{\nu+1}
                     \int_{0}^{\pi}dX \, e^{-p X}\sin^{\nu} X
\nonumber\\
&=& \frac{\xi^2}{2} \left( \frac{2w}{\pi\xi} \right)^{\nu+1} 
Z(\pi),  \label{eq20}
\end{eqnarray}
where $p=\nu I/I_0$ and $\nu=\epsilon_0/T$ and
$\epsilon_0(t)$ and $I_0(t)$ are assumed  weakly time-dependent in the potential  of Eq.\,(\ref{energy}).
The integral $Z(\pi)$ has an exact solution with an asymptotic expansion for $\nu\gg 1$: \cite{Integrals1, Integrals2}
\begin{eqnarray}
&&Z(\pi) =
\frac{\pi \exp(-\pi p/2)\Gamma(\nu+1)}{ {2^{\nu}} |\Gamma(1+\nu/2+ip/2)|^2} \nonumber \\
&&\sim
\sqrt{ \frac{2\pi}{\nu} } \left[ 1 + \left(\frac{p}{\nu+2}\right)^2 \right]^{-\frac{\nu+1}{2}} 
\!\!\! e^{-p \,{\rm tan}^{-1} \frac{\nu+2}{p}}
\end{eqnarray}
with the Gamma function $\Gamma(z)$.
Finally, for $I$ close to $I_{c0}$ we find the total vortex and antivortex-crossing rate 
$R_{tot} = R_v + R_{av} = 2 R_v$ at $H=0$ in a belt of size $w\times w$:
\begin{eqnarray}
 R_v(I,T) &\approx& \Omega(I, T, \nu(t)) \left[\frac{I}{I_{c0}(t)}\right]^{\nu(t)+1} , 
\label{rate0} \\
 \Omega(I, T, \nu) &\approx& \frac{4\pi T c^2 R_{\rm eff}}{ e \Phi_0^2} \sqrt{ \frac{\nu}{2\pi} }
\nonumber\\ 
&& \times e^{-\frac{\nu}{3}(\pi e \xi I_{c0}(t)/2 w I)^2} , 
\\
 R_{\rm eff} &=& \frac{R_s R_{\square}}{R_s+2\pi(\xi/w)^2 R_{\square}}, 
\label{rate}
\end{eqnarray}
The solution for the  parameters changing with time on the characteristic time scale  
$\tau$ the adiabatic approach is valid if the thermalization time is much shorter than $\tau$, i.e., if
\begin{equation} 
\eta_{\rm eff}/
\frac{\partial^2U}{\partial x_v^2}|_{x_v=x_s}\approx \frac{e^2{\tilde\Lambda}}{c^2R_{\rm eff}}\ll \tau .
\end{equation} 
This condition is satisfied if $\tau \ll 0.1$ ps for $R_{\rm eff} = 400 \, \Omega$ and
$\tilde\Lambda = 60\, \mu m$.

For currents $I \simeq I_{c0}$ the exponential factor on the right-hand-side of the attempt frequency $\Omega$ is of order $\sim 1/e$ for materials parameters of interest
($w/\xi\sim 25$ and $\nu\sim 110$). Further assuming $R_{\rm eff}=400$ $\Omega$, $T=5$ K, and $I=I_{c0}$, 
we estimate $\Omega \approx  4.4\times10^{10}$ s$^{-1}$.
However, one should not put too much weight on the estimate of the `pre-exponential' factor $\Omega$ compared to}
$(I/I_{c0})^{\nu+1}$, because
the approach ``vortex as a particle'' was used and such an approach is only approximate near the strip's edges, resulting 
in possible corrections of order unity.

It is worthwhile to point out that our expression for the rate $R_v$ is smaller by a large factor $2^{\nu} (\nu-1)w/\xi = 3.5\times10^{36}$ compared to the voltage-current result derived in Ref.~\onlinecite{gv}.
There are three reasons for this significant discrepancy: 
(i) the factor $2^\nu$ originates from the logarithmic difference in the vortex potential $U$ in Eq.~(\ref{energy}), 
(ii) the factor $(\nu-1)$ is due to the difference in boundary conditions (periodic vs.\ non-periodic) for a finite-size geometry, and 
(iii) the factor $w/\xi$ stems from the fact that vortices are uncorrelated (non-interacting) at distances 
larger than $w$ along the $y$-axis 
and not $\xi$, as was assumed in Ref.~\onlinecite{gv}. 

Another noteworthy result of this work is that a shunt suppresses the  rate because the effective vortex viscosity increases due to the additional dissipation in the combined system of superconducting strip and normal shunt.
Finally, we note that at large $\nu$ the condition of small probability current density $S(t)$ is fulfilled for practically all currents of interest. Thus justifying our original assumptions for solving the Smoluchowski equation.

\subsubsection{Field-dependent rate}

To evaluate the magnetic field effect on the rates, we replace $I$ with $I_+$ of Eq.\,(\ref{barrier})
as discussed above. Then, one can see that  
 the vortex-crossing rate increases with field according to
\begin{eqnarray}
R_v(I,H) \equiv R_v(I_{+}) \approx R_v(I,0)\left(1+\frac{I_{c0} H}{I H^*}\right)^{\nu+1}\,.
\end{eqnarray}
For $I_{c0} H\ll I H^*$ and in the large limit for the exponent, $\nu\gg 1$, we obtain for the vortex-crossing rate
\begin{eqnarray}
R_v(I,H)  \approx R_v(I,0) e^{H/H_1}\,,\ H_1=\frac{H^* I}{(\nu+1) I_{c0} }\,.
\label{eq:H_1}
\end{eqnarray}
Next, we consider the crossing rate for antivortices entering from the right edge $x=w$ and moving toward $x=0$. The zero-field potential barrier for antivortex is a reflection of that for a vortex in the strip middle, i.e., the zero-field rates for vortices entering at $x=0$ and for antivortices starting from $x=w$ are the same. However, the field contribution $-M_aH$ to the antivortex energy  has opposite sign to that of a vortex. It is straightforward to obtain the corresponding barrier
\begin{equation}
\frac{\epsilon_b}{\epsilon_0}=
\ln \frac{I_{c0}}{I_{-}} \, ,\qquad I_{-}=I-I_{c0}\frac{H}{H^*}\,.
\label{barrier-}
\end{equation}
We then obtain the rate for antivortices after similar calculations
\begin{eqnarray}
R_{av}(I,H)  \approx R_v(I,0) e^{-H/H_1} .
\end{eqnarray}
The total crossing rate is the sum of vortex and antivortex crossings
\begin{equation}
R_{tot}(I,H)=2R_v(I,0)\cosh(H/H_1)\,. 
\label{total}
\end{equation}

\subsection{Characteristic magnetic fields}

Next, we estimate all characteristic fields for typical film and materials parameters of the SNSPD.
We consider a film with width $w=100$ nm, thickness $d=6$ nm, coherence length $\xi=4$ nm, and renormalized
Pearl length $\tilde{\Lambda}=60$ $\mu$m at temperature $T=5.5$ K 
with the parameter  $\nu_0\approx 110$, see Ref.~\onlinecite{bgbk}.
In this case, the characteristic magnetic field for describing the potential change is $H_0 = 0.1$  T, see Eq.~(\ref{energy}).
The field for vortex entry  is $H_{c1}=0.36$ T, see Eq.~(\ref{eq:Hc1}).
The field scale for the critical current $I_c(H)$ is $H^* \approx 0.6$ T, see Eq.~(\ref{eq:H*}),
and for  the vortex-crossing rate is
$H_1\approx 4.1$ mT at $I= 0.75I_{c0}$, see Eq.~(\ref{eq:H_1}).

\section{Transition S$\Rightarrow$N caused by vortex crossing}

It was argued in Ref.~\onlinecite{bgbk} that a single vortex, crossing at high currents, can destroy the superconductivity 
inside the belt, which covers the entire width of strip. If so, the vortex crossing results in the switching of the current 
from superconducting strip to shunt 
in SNSPD and the destruction of superconductivity in the whole strip, when the shunt is absent or insufficient. 
Here, we show that it occurs in strips with the width smaller than some critical $w_c$. We will estimate $w_c$ and the 
current $I^*$ above which it happens.

A crossing vortex releases the energy $\Phi_0I/c$ as
a cloud of quasiparticles is squeezed out of the moving vortex core as described by Larkin and Ovchinnikov. \cite{LO}
These quasiparticles diffuse from the vortex path during  crossing time $\tau_0\approx w^2\Phi_0/(2\pi\xi^2cR_{\square}I)$ 
of the order of several pico-seconds for $w=100$ nm and 
bias currents of the order   $I_c$. This estimate was obtained in Ref.~\onlinecite{bgbk} assuming that the supercurrent is not affected by 
vortex crossings and thus the vortex moves under the effect of almost constant Lorentz force. 
During the  time $\tau_0$ of the order of several pico-seconds, the energy transfer  to
phonons and substrate is negligible.\cite{Semenov}  
Thus, when the vortex exits, the quasiparticle cloud has approximately the form of 
  a belt of  width  
 $\ell(I)\approx (D\tau_0)^{1/2}$, where $D$ is the diffusion coefficient. 
We estimate the width as
\begin{equation}
\ell(I)=\xi\left(\frac{2e\pi \tilde\Lambda w D}{c^2 R_{\square}\xi^3}\frac{I_c}{I}\right)^{1/2} 
\approx 4.29 \xi\left(\frac{I_c}{I}\right)^{1/2} \,,
\end{equation}
where $R_\square=400\, \Omega$ is the film sheet resistance,  $D \approx 0.46$ cm$^2$/s is the diffusion coefficient, and
$\tilde\Lambda= 60\, \mu$m, $\xi=4$ nm, $w=100$ nm.
We estimated the diffusion coefficient from $D=\pi^2 k_B^2/(3 q^2 \gamma d R_\square)$, where $q$ is the charge of the electron, $k_B$ is Boltzmann's constant, and $\gamma=220$ J/m$^3$K$^2$ is the Sommerfeld coefficient.
 The  energy released by a vortex crossing, $\Phi_0I/c$,  suffices to destroy superconductivity in
the area of the quasiparticle cloud (belt) at currents $I>I^*$,  where $I^*$ in zero-field is defined by the energy balance
\begin{equation}
\frac{\Phi_0 }{c}I^*=\frac{H_c^2}{8\pi}\ell(I^*)wd.
\label{dark_count_I}
\end{equation}
Our estimate for the hot area results in a lower bound for the threshold current $I^*$, where the
thermodynamic critical field of the superconducting condensate is defined by
$H_c^2 = \Phi_0^2/(4\pi^2\tilde\Lambda d \xi^2)$.
We attain for the critical threshold current
\begin{equation}
\frac{I^*}{I_c} \approx \frac{e\ell(I^*)}{8\pi\xi}
= \left( \frac{e^{3} \tilde{\Lambda} w D}{32\pi c^2 R_{\square} \xi^3} \right)^{1/3} \approx 0.60.
\end{equation}
To have $I^*<I_c$ one needs to satisfy the condition for the critical width of the strip 
with materials parameters used above
\begin{equation}
w_c < 1.745\cdot10^6\xi^2/\tilde\Lambda  , 
\label{condition}\end{equation}
otherwise  a single-vortex crossing does not result in the formation of normal belt across the entire strip width and the 
superconducting state is stable with respect to single-vortex crossing.  
For  films studied so far, we have $\tilde\Lambda/\xi \approx 1.5\cdot 10^4$ and thus the condition Eq.~(\ref{condition}) 
is fulfilled for strips of widths up
to $w_c \approx 116\xi$.  In this case, the
vortex crossing destroys  the metastable state at  currents $I>I^*$ without shunt and 
produces a voltage pulse  in the combined system with shunt. 
Our estimate for $I^* \agt 0.6 I_c$  is rather crude.
It neglects the possible creation of a normal region behind the moving vortex and the corresponding 
change of the supercurrent density in front of it.

\section{Vortex-assisted photon counts}\label{vortex-assisted-counts}

Absorption of a single photon in a thin and narrow superconducting strip 
results in      
a cloud of quasiparticles with energies above the superconducting gap.
First, this cloud of excitations grows in number of quasiparticles 
and in size due to avalanche processes and diffusion,\cite{Brorson, Semenov} 
but in time the  cloud diminishes and finally vanishes
due to relaxation processes. We assume that 
in sufficiently narrow strips, $w\lesssim 100 $\,nm, 
(a)    the cloud covers the entire width $w$ of strip  when it reaches its
maximum size,
(b) the   quasiparticle density in the cloud (hot belt) is close to uniform, and 
(c) quasiparticles suppress the superconducting order parameter inside the hot belt, 
but {\it their density is not sufficient to convert the hot belt  to the normal state}. 
Thus the superconducting condensation energy density in  the hot belt, $H_{ch}^2/8\pi$,  
satisfies the inequalities $0<H_{ch}^2/8\pi<H_{c}^2/8\pi$. 
Further, we assume that quasiparticles in the hot belt thermalize exponentially with relaxation time 
$\tau$ of the order of 40 ps for film thickness considered, see Refs.~\onlinecite{Ilin2000, Beck2011}. 
It follows that the vortex-crossing rate via hot belt (denoted by subscript $h$), $R_{vh}(I)$, is enhanced in comparison with that for dark counts,
because the parameter $\nu_h=\epsilon_{0h}/T\sim H_{ch}^2/(8\pi T)$ is reduced in comparison with $\nu_0$.
In the framework of assumptions (a) and (b), the crossing rate inside the hot belt is given by   Eq.~(\ref{total}), 
however, with renormalized parameters $\epsilon_{0h}<\epsilon_0$,   diminished 
exponent $\nu_h=\epsilon_{0h}/T<\nu_0$, and  reduced critical current   
$I_{ch}=I_{c0}( \nu_h/\nu_0)<I_{c0}$.

\subsection{Current dependence of assisted counts}

\begin{figure}[t]
\includegraphics[width=80mm]{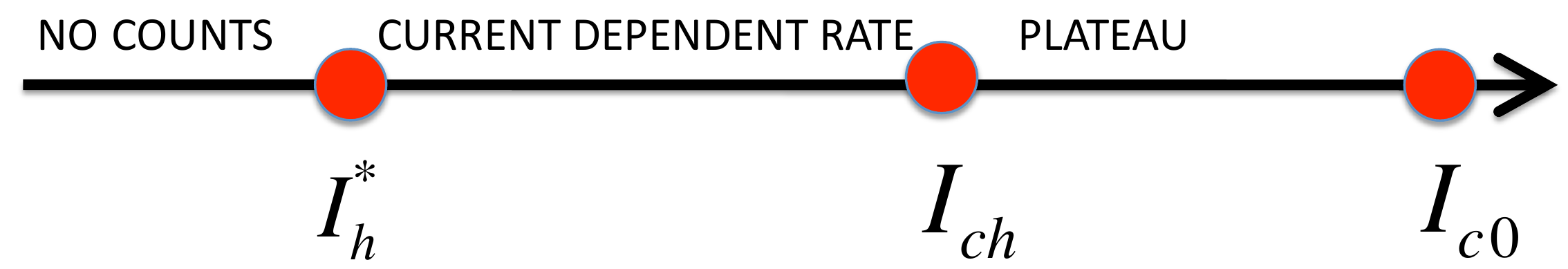}
\caption{(Color online)
Three characteristic current regimes and three characteristic behaviors for photon count rates.
}
\label{fig:current}
\end{figure}

There are three superconducting current-biased regions to be considered, which are separated by three characteristic currents $I_h^* < I_{ch} < I_{c0}$,
see Fig.~{\ref{fig:current}}.
For currents below the {\it hot threshold}, $I_h^*$, counts are absent, because vortex crossings do not result in the formation of a normal-state belt.
$I_h^*$ is determined through the energy balance for destroying the superconducting condensate in the hot belt similar to that described in    Eq.~(\ref{dark_count_I}) for dark counts, however, with the  thermodynamic field $H_{ch}\sim H_c \sqrt{\nu_h/\nu_0} < H_{c}$:
\begin{equation}
\frac{\Phi_0I_h^*}{c} = \frac{H_{ch}^2}{8\pi}\ell(I_h^*)wd ,
\end{equation}
and hence $I_h^* \approx 0.6 I_{ch}$.
Between $I_h^*$ and $I_{ch}$, the vortex-assisted photon counts are strongly  current dependent  with a precipitous drop below $I_{ch}$. 
At high currents, $I_{ch} < I < I_{c0}$, photon count rates saturate at a plateau.
In the region  $I_h^* < I < I^* \approx 0.6 I_{c0}$, vortex crossings via the hot belt destroy the superconducting state inside the belt
during the relaxation time $\tau$   of the hot spot. During this time, the belt is in the normal state, which causes 
 the ``vortex-assisted'' photon count.
 
\begin{figure}[tbh]
\includegraphics[width=80mm]{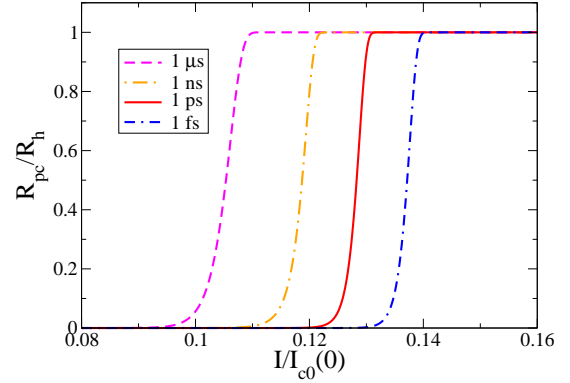}
\includegraphics[width=80mm]{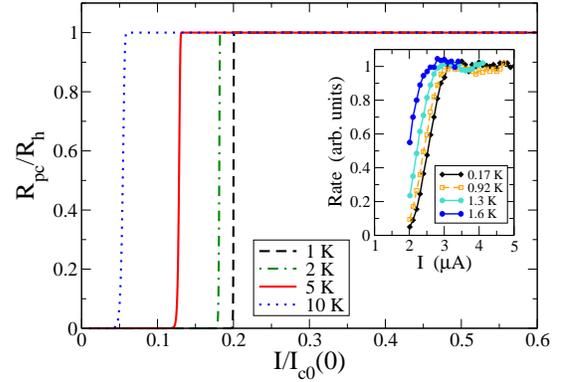}
\caption{(Color online)
The vortex-assisted photon count rate $R_{pc}$ vs.\ bias current given in Eq.~(\ref{ph}).
Top panel: Rates for several values of hot belt lifetime $\tau$ at $T=5$ K.
Bottom panel: Rates for several temperatures at $\tau=1$ ps. Other parameters for NbN are
$\nu_h=40$ and $\nu_0=110$ at T=5 K, $T_c=15$ K, $R_p=R_h=10^6 \, {\rm s}^{-1}$, $R$= 400 $\Omega$, $w=100$ nm, 
and $\xi_0=4$ nm. Inset: Current dependence of
photon count rates in a-W$_x$Si$_{1-x}$ for different bath temperatures from Ref.~\onlinecite{Baek}. Note, that 
film temperature is generally higher than indicated due to heat released by photons and corresponding vortex crossings.
}
\label{fig:Rpc}
\end{figure}

If $I_{ch} < I < I_{c0}$, the rate of single-photon counts,
$R_{pc}$, is the same as the rate for hot spot formation $R_h$, because the barrier for 
vortex crossing is now absent. For hot counts $R_h$, we have $R_h=\eta_h R_p$, where $R_p$ is the photon rate and 
$\eta_h$ is the quantum efficiency of hot spot formation caused by a single photon. \cite{Semenov, Baek}

Next, we consider the region $I_h^*<I<I_{ch}$, where the photon count rate depends on the bias current. This  
dependence has been observed in NbN strips with   $d=5$\,nm and  
 $w=97$\,nm for photons with 500\,nm wavelength,\cite{Semlast}
as well as in a-W$_x$Si$_{1-x}$ SNSDPs ($d=4.5$\,nm, $w=150$\,nm) for a wide range of photons of 672-1850\,nm.\cite{Baek}
In this current regime,  the photon count rate is $R_{pc}=R_h{\cal P}(R_h^{-1})$, 
where ${\cal P}(t)$ is the probability of vortex crossing before time 
$t$;  $t$ is counted from the moment of the hot spot formation. 
For $R_{pc}$, we take ${\cal P}(t)$ at
$t=R_h^{-1}$, 
because a given hot 
spot does not contribute to photon counts if the vortex crossing does not occur before   formation of another
hot spot by the next incident photon. The probability ${\cal P}(t)$ satisfies  the rate equation
\begin{equation}
\frac{d{\cal P}}{dt}=R_{vh}(I, t)\big(1-{\cal P} \big)\,,
\end{equation}
which yields
\begin{equation}
{\cal P}(t)= 1- \exp\left[-\int_0^{t}dt' R_{vh}(I, t')\right].
\label{P-eq}
\end{equation}
Due to the thermal relaxation of the hot spot the
vortex-crossing rate via the hot spot changes with time, because  
$I_{ch}(t)$ and $\nu_h(t)$ diminish with $t$. We will account only for time dependence of $\nu$ as the crossing rate is much more 
sensitive to $\nu$ than to $I_{ch}$. 
For $\nu(t)$ we have
\begin{equation}
\nu(t)=\nu_0+1+(\nu_h-\nu_0)\exp(-t/\tau),
\label{nu_t}
\end{equation}
where $\nu_0$ is defined in the   absence of the hot spot, that is, it is the same as for dark counts. 
Defining the integral in ${\cal P}(R_h^{-1})$ in Eq.~(\ref{P-eq}) as
\begin{eqnarray}
{\cal R}(I,\nu_h) &\equiv&
\int_0^{R_h^{-1}}dtR_{vh}(I,t)=\int_{\nu_h+1}^{\nu(R_h^{-1})}d\nu \frac{dt}{d\nu}R_{vh}(I,\nu),  \nonumber\\
&&\frac{dt}{d\nu}=\frac{\tau}{\nu_0+1-\nu},
\label{phg}
\end{eqnarray}
we obtain 
\begin{equation}
{\cal R}(I,\nu_h)=\frac{ \Omega(I, T, \nu_h) \tau }{ \sqrt{\nu_h} } \int_{\nu_h+1}^{\nu_1} \frac{d\nu\,\nu^{1/2}}{\nu_0+1-\nu}
\left(\frac{I}{I_{ch}}\right)^{\nu},
\end{equation}
where $\nu_1 \equiv \nu(R_h^{-1})$ is defined by Eq.~(\ref{nu_t}). The integrand increases rapidly as $\nu$ changes from  
 $\nu_h$ to a  larger value $\nu_1 \approx \nu_0+1$ when $R_h\tau \ll 1$.
Hence, the main contribution comes from the 
region near $\nu_h$. 
We define $\nu=\nu_h+1+\alpha$ and
expand the integrand in $\alpha$. Integrating over $\alpha$ we find 
\begin{equation}
{\cal R}\approx\frac{\tau R_v(I,\nu_h)}{\nu_0-\nu_h}\,\,\frac{1-(I_{ch}/I)^{\nu_h-\nu_1-1}}{\ln(I_{ch}/I)},
\end{equation}where
\begin{equation}
R_v(I,\nu_h) \approx \Omega(I,T,\nu_h)
\left[ \frac{I}{I_{ch}} \right]^{\nu_h+1}.
\end{equation}
Finally, we attain the vortex-assisted photon count rate
\begin{equation}
R_{pc}(I,\nu_h)=R_h[1-\exp(-{\cal R}(I,\nu_h)].
\label{ph}
\end{equation}
The   measured count rate includes  dark counts in the remaining "cold" part of the strip, which must be added to
the right-hand-side of Eq.\,(\ref{ph}). Roughly, if $R_{pc} \gg R_v (I,\nu_0)L/w$ one can neglect 
the contribution of dark counts from the "cold"
section of the strip.

At high currents  $I>I_{ch}$,  
we obtain $R_{pc}\approx R_p$, that is, the photon count rate
has a plateau   as a function 
of $I$. Here, experimental data on photon rates  provide information on the 
rate of hot spot formation by incident photons, $R_{p}$.
For lower currents  in the interval $I_{ch}>I>I_{h}^*$,
the rate of photon 
counts shows a power-law current dependence, 
\begin{equation}
R_{pc}(I)\approx R_h{\cal R}\approx R_h \tau R_v(I,\nu_h)/(\nu_0-\nu_h) \propto I^{\nu_h+1}.
\label{eq_Rpc}
\end{equation}
The low-current part of the curve $\ln R_{pc}$ vs $\ln I$ 
allows us to extract   $\nu_h$ characterizing the suppression of 
superconductivity in  the hot spot. 
A sharp crossover from high to low current behavior occurs at   current values    close to the critical current $I_{ch}$ 
due to high values of $\nu_h$. This crossover depends weakly on $\tau$, as one sees from the top panel in Fig.~\ref{fig:Rpc}. 
Furthermore, at bias currents below 
$I_h^*$ photons do not lead to   formation of normal belts and thus are not counted.

To prove that vortex crossings are involved in photon counts at $I>I_h^*$ one may use the magnetic field 
which enhances the vortex-crossing rate along with photon counts at low currents. 
For $H\ll H^*$, and accounting for both vortex and antivortex crossings in the hot belt,
this enhancement is described by the relation
\begin{equation}
R_{pc}(I,H)=R_h\Big( 1-\exp[ {-2{\cal R}(I,\nu_h)\cosh( H/H_{1h})}] \Big) .
\label{pc}
\end{equation}
with field scale $H_{1h} = H_1 (\nu_0+1)/(\nu_h+1)$.
The field effect    is most pronounced at low currents when ${\cal R}(I,\nu_h) \ll 1$.
Since the magnetic field  affects vortex crossings, but not the formation of hot spots by photons, 
the dependence of vortex-assisted counts $R_{pc}$ on $H$ will unambiguously confirm 
the involvement of vortices  in photon counting.

For wider strips, when the hot spots of excited quasiparticles do  not cover all of the strip  width, 
  the hot spot  diminishes locally the effective width of the superconducting strip and   
 roughly one can replace 
$w\to\tilde w=w-D$ in Eqns.~(\ref{rate0}) and (\ref{rate}), where $D$ is the hot spot diameter. 
Thus we can use Eq.\,(\ref{pc}) with the renormalized width $\tilde w$ and the same $\nu_0$ as for   dark counts. 
It is worth to remember that the vortex  crossing rate depends strongly on $w$, 
$R_v(I,\nu_0)\propto w^{-\nu_0-1}$, with $\nu_0\approx 110$.
Hence even a small decrease
in $w$ enhances dramatically the vortex-crossing rate due to the formation of  hot spots. 
Clearly, for a more rigorous treatment one needs to consider  vortex crossings  for a nonuniform superconducting 
order parameter and  nonuniform currents.
However, our main result will stay:   the magnetic field renormalizes the bias current as described above and results in the enhancement of photon count rates.

\subsection{Comparison with experiments}

In the framework of our model, the data presented in Fig.~4 
of Ref.~\onlinecite{Semlast} for nanowires with $w=97$\,nm yield an estimated  
$\nu_h\approx 40$ at 6 K. On the other hand, for strips of  similar width and thickness, $\nu_0$ 
for the dark count rate at 5.5 K 
was estimated as $\nu_0 \approx 110$, see Ref.~\onlinecite{bgbk}, where the data reported 
in Ref.~\onlinecite{Bartolf} were used. 
Hence, the barrier suppression by the field  is  strong; still, more measurements to extract $\nu_0$ and $\nu_h$ 
are needed, especially for narrow 
strips for which our model of uniform hot spots across the whole strip width is applicable.

The part of the curve $R_{pc}(I)$ at intermediate currents provides information on the 
relaxation time $\tau$. The dependence of $R_{pc}(I)$ for $\nu_h=40$ and $\nu_0=110$ at 5 K for several values of $\tau$ 
and temperatures is presented in Fig.\,\ref{fig:Rpc}. We model the temperature dependent parameters
$I_{ch}(T) = I_{ch}(0) \varepsilon_T^3$, 
$\xi(T)=\xi_0/\varepsilon_T$, and 
$\nu_i(T) = \nu_i(0) \varepsilon_T^2/T$ with $\varepsilon_T = (1-T/T_c)^{1/2}$.
Recently, similar behavior has been reported by the NIST group \cite{Baek} for the current 
dependence of the photon count rate in  a-W$_x$Si$_{1-x}$ strip with $T_c\approx 3.8$ K. The data for $R_{pc}/R_p$ are shown in the inset of the lower panel of Fig.\,\ref{fig:Rpc}. 

Both calculated and observed current dependences of photon count rates have a plateau at high currents and a sharp drop below the current associated with $I_{ch}$. 
Note that the superconducting condensation energy in  the hot belt is proportional to $\Phi_0I_{ch}/c$.
Therefore one can anticipate that the condensation energy in the belt drops with increased photon energy.
The wavelength dependence of the count rate in Fig.~2(b) in Ref.~\onlinecite{Baek} is in agreement with such an expectation.

\section{Discussion and conclusions}

In a somewhat oversimplified picture presented in the literature, 
the hot spot formation results directly in a voltage pulse. 
The argument goes like that: The
hot spot becomes normal and  the transport supercurrent  should redistribute in   the remaining superconducting regions. There, it 
exceeds the   critical value  and  transforms this area 
 to the normal state as well. 
Because of the large resistance of the normal belt created, it induces the current redistribution to the shunt resistor and a voltage pulse is detected. 
To quantify this scenario,  we presented in this work a vortex-based phenomenological model for the transition from the
current-carrying metastable superconducting 
state  to the current-carrying normal state. Two important points in our scenario are the metastability of the current-carrying superconducting state 
with respect to vortex crossings and the energy balance of the transition process  
 driven by such a crossing. 

One-dimensional wires (nanowires) with $w\lesssim \xi$ and two-dimensional wires 
 with $w\gg \xi$, considered in this work, have many features in common.
As we have already shown in Ref.~\onlinecite{bgbk}, phase slips in 1D wires \cite{Amb,McC} play the same role as vortex crossings in 2D wires at low temperatures considered above.
Each phase slip in nanowires releases the energy
$\Phi_0I/c$.  The critical current in nanowires,
$I_{c0}=c\Phi_0\xi/(12 \sqrt{3} \pi^2 \lambda^2)$, 
  is only slightly larger than for the strip.
For currents $I=I^* \gtrsim 0.3 I_{c0}$,
the energy released by phase slips is
sufficient to destroy superconductivity in a segment of the nanowire of
cross-section $\pi\xi^2$ with length $\sim 1.2\xi$.
 Eventually, the whole wire becomes normal due to Joule heating
{\it if the shunt is absent}. Therefore, in this case there is no {\it dc} voltage to be considered in the superconducting state.
Similarly, at currents well below $\approx 0.3 I_{c0}$ the probability of 
phase slips is negligible and no {\it dc} voltage may be observed.
On the other hand, in the presence of a shunt, voltage pulses will be detected in full
analogy with dark counts in shunted strips.
Note, that the shunt resistance is even more important in phase slip rates than for strips,
because the small factor $\xi^2/w^2$ in Eq.~(\ref{rate}) of our model becomes  of order unity in nanowires
and strongly renormalizes the viscosity.

Interestingly, the proposed SNSPD mechanism is similar to that of the bubble chamber used to detect charged 
elementary particles in particle physics.\cite{Glaser} Inside the chamber, a superheated liquid  is prepared in a metastable state. 
On entry of a charged particle, the ionization track is formed around which the liquid vaporizes by forming mesoscopic bubbles. 
The bubble density around  the track is proportional to the particle energy loss. In  SNSPD too,  the transition from the metastable  to  stable states is used for the detection of a single photon. 

We summarize the calculations within the phenomenological model of vortex-assisted photon counts in narrow and thin SNSPDs by listing the main results:

(a) We estimated the critical width $w_c$ of the superconducting strip below which single-vortex crossings create a
normal belt across the entire strip width resulting in dark counts in the SNSPD and transition to the normal state in the absence of a shunt.

(b) We have shown that the magnetic field in combination with a bias current enhances the vortex 
crossing rate. At currents $I\gg I_0$, close to $I_c$, the effect of the magnetic field is equivalent  to an increase of the bias 
current by a factor $(1+H/H^*)$, where $H^*\approx 0.6$\,T  in 6\,nm thick  
and  100\,nm wide films. The corresponding characteristic field affects the vortex-crossing rate (dark counts), 
because the effective current is much smaller, roughly by  a factor $1/\nu_0\ll 1$. For films of similar geometry, fields $H\ll H^*$, 
and for currents $I=0.75 \, I_c(0)$
the vortex-crossing rate is renormalized by a  factor 
$\cosh(H/H_1)$, where $H_1\sim H^*/\nu_0  \approx 4$\,mT.

(c) We have derived the photon count rates   that require  a vortex crossing to 
transform the photon-created hot spot  to a normal-state belt across the 
 strip's width. A single photon incident diminishes the critical current in  the hot spot eliminating the barrier for 
vortex crossing at currents $I>I_{ch}$ (within the plateau region of the count rate). 
This results in a photon count when  
 the photon energy  is too small to create
a normal-state belt. At currents below $I_{ch}$ a reduced barrier still exists and the photon count rate depends strongly on the bias current.

(d) We also  argued, based on available   data, \cite{Semlast,Baek} that 
vortex crossings are the origin of photon counts in the SNSPD.
This vortex-assisted mechanism  may
 be verified by the application of magnetic fields, which enhances  the vortex 
crossing rates, but does  not affect the creation of hot spots by photons. 
We predict  that already weak magnetic fields of order $H_1$ should cause  the
  enhancement of photon counts, in particular at low bias currents.

Finally, our calculations show  that the vortex-assisted photon count rate is characterized by a plateau at high currents  with a tail toward low currents.
This feature suggests that an optimum current exists, which  minimizes the effect of dark counts, while only weakly diminishing
photon counts. We propose to operate SNSPD devices above $I_{ch}$, the low-current edge of 
the plateau,  where photon-induced counts are still  effective, whereas the dark count rate is strongly suppressed.

\acknowledgments
We like to thank A. Gurevich, Shizeng Lin, A. Schilling, A. Engel, 
M. Siegel, I. Martin, R.D. McDonald, O.A. Valenzuela, F. Ronning,
M.W. Rabin, and  N.R. Weisse-Bernstein for discussions. 
Work at the Los Alamos National Laboratory was performed under the auspices of the
U.S.\ DOE contract No.~DE-AC52-06NA25396.
Work at the Ames Laboratory (VK) was supported by the U.S.\ DOE, Office of Basic
Energy Sciences, Division of Materials Sciences and Engineering under
Contract No.~DE-AC02-07CH11358.

\end{document}